\documentclass[11pt,letterpaper]{article}
\usepackage[letterpaper,total={6.5in, 9in}]{geometry}

\usepackage{changepage}

\usepackage[utf8]{inputenc}

\usepackage{textcomp,marvosym}

\usepackage{fixltx2e}

\usepackage{amsmath,amssymb}

\usepackage{cite}

\usepackage{nameref,hyperref}

\usepackage{microtype}
\DisableLigatures[f]{encoding = *, family = * }

\usepackage[aboveskip=1pt,labelfont=bf,labelsep=period,justification=raggedright,singlelinecheck=off]{caption}

\usepackage{relsize}
\usepackage{graphicx}
\usepackage{color}
\usepackage{amssymb}
\usepackage{mathrsfs}
\usepackage{mathtools}
\usepackage{authblk}

\title{Designing robust watermark barcodes for multiplex long-read sequencing}
\author[1,2,*]{Joaqu\'in Ezpeleta}
\author[1]{Flavia J. Krsticevic}
\author[1,2]{Pilar Bulacio}
\author[1,2]{Elizabeth Tapia}
\affil[1]{Centro Internacional Franco Argentino de Ciencias de la Informaci\'on y de Sistemas, Rosario, Argentina; \textsuperscript{2}Facultad de Ciencias Exactas, Ingenier\'ia y Agrimensura, Universidad Nacional de Rosario, Rosario, Argentina}
\affil[*]{ezpeleta@cifasis-conicet.gov.ar}
\date{}

\let\OLDthebibliography\thebibliography
\renewcommand\thebibliography[1]{
  \OLDthebibliography{#1}
  \setlength{\parskip}{0pt}
  \setlength{\itemsep}{2pt plus 0.3ex}
}

\makeatletter
\renewcommand{\@biblabel}[1]{#1.}
\makeatother

\begin{document}
\maketitle
\vspace{-1cm}
\paragraph*{Abstract}
A method for designing sequencing barcodes that can withstand a large number of insertion, deletion and substitution errors and are suitable for use in multiplex single-molecule real-time sequencing is presented. The manuscript focuses on the design of barcodes for full-length single-pass reads, impaired by challenging error rates in the order of 11\%. To the authors' knowledge, this is the first method to specifically address this problem without requiring upstream quality improvement. The proposed barcodes can multiplex hundreds or thousands of samples while achieving sample misassignment probabilities as low as $10^{-7}$, and are designed to be compatible with chemical constraints imposed by the sequencing process. Software for constructing watermark barcode sets and demultiplexing barcoded reads, together with example sets of barcodes and synthetic barcoded reads, are freely available at \url{www.cifasis-conicet.gov.ar/ezpeleta/NS-watermark}.
\section*{Introduction}
Single-Molecule Real-Time (SMRT) sequencing, with average read lengths of {\raise.17ex\hbox{$\scriptstyle\sim$}}10 kbp \cite{numberOfPasses}, 
is poised to remarkably simplify genome assembly and targeted sequencing in many applications \cite{Koren2015, Guo2015, substitutionErrorRate}. In this new era, DNA reads are considerably longer, but unfortunately corrupted by unusually high rates of sequencing errors. For SMRT sequencing, error rates of {\raise.17ex\hbox{$\scriptstyle\sim$}}11\% \cite{totalErrorRate}, with a predominance of insertions/deletions (indels) and only {\raise.17ex\hbox{$\scriptstyle\sim$}}1\% substitution errors\cite{substitutionErrorRate}, must be considered. Fortunately, in any information transmission process affected by noise \textemdash in this case SMRT sequencing\textemdash ~errors can be corrected by adding enough redundancy to transmitted information \textemdash in this case DNA sequences\textemdash ~\cite{gall1968}. The simplest way to add redundancy is to transmit multiple copies of the information in the hope that it will be possible to recover the original data through some form of consensus. This is, for example, the rationale behind genome oversampling or coverage, which is used across sequencing technologies to obtain virtually error-free sequences from noisy reads. We will see, however, that this approach cannot be directly applied to parallel multiplex \cite{Hamady2008} SMRT sequencing without sacrificing the much desirable read length.

Multiplex sequencing relies on the use of short oligonucleotides, known as barcodes, to tag DNA fragments belonging to different samples, which provides a means for translating the massive throughput of next-generation sequencing (NGS) technologies into reduced sequencing costs. Barcodes are sequenced and identified to assign each read to a sample, which is known as demultiplexing. A variant of SMRT sequencing called Circular Consensus Sequencing (CCS), which provides more accurate reads, is generally advised for SMRT multiplex applications \cite{CCS}. In this variant, SMRT reads of improved quality are generated from intra-molecular consensus over sub-reads obtained from multiple ($\geq 2 \times$) polymerase passes along a circularized sequencing template, which follows the repetition approach described above. With CCS, however, the length of native reads is reduced by a factor equal to the number of polymerase passes. To attain Illumina-level quality values, about five passes are needed and the effective read length drops to a few kilobases \cite{numberOfPasses}.

For the longer read length afforded by SMRT sequencing to be fully leveraged, multiplexing would need to work directly with single-pass reads, known as Continuous Long Reads~(CLRs). In this scenario, all error correction would depend on redundancy embedded in the barcodes themselves. However, it has been noted that current barcodes for SMRT-CCS reads, with lengths below 20~nt, are not sufficiently robust and that longer barcodes would be needed for this purpose \cite{longerBarcodes}. This comes as no surprise, since most existing SMRT-CCS barcodes are obtained by imposing a minimum edit distance constraint to sets of random sequences \cite{Buschmann}, a design method known to scale poorly with increasing barcode length \cite{masek1980faster}. To overcome the limitations of random barcodes in general multiplex SMRT sequencing applications, systematic barcodes can be alternatively explored.

Recently, watermark barcodes, a class of systematic barcodes able to deal with sequencing indels and substitutions, have been proposed \cite{Kracht}. These are inspired in the design of watermark error correcting codes \cite{watermarks}, originally developed to deal with synchronization and substitution errors in digital communications. In these applications, synchronization errors are modeled as the random insertion and deletion of symbols \cite{leven1966}, and are thus assimilable to sequencing indels.

Briefly, watermark codes consist of an information-containing carrier sequence imprinted with an arbitrary but fixed sequence of equal length, known as watermark. In the original formulation of watermark codes, the carrier sequence is sparse, meaning it contains a majority of null or ``zero'' symbols (i.e. symbols which, when imprinted with \textemdash or added to\textemdash ~another, will not modify it, much like zero in regular arithmetic). To obtain this carrier sequence, an information message is protected by a tandem of two error correcting codes, known as outer and inner code. Both of these embed redundancy in the form of additional symbols, so that the resulting sequence is considerably longer than the original message. For example, one of a set of 48 DNA samples, which in principle requires only three bases ($4^{3}=64 \ge 48$), might be represented by a carrier of, say, 16 symbols, most of the null type. When the watermark is imprinted over the carrier, the sparse constraint will imply that the resulting sequence will match the watermark at most positions. Since the watermark is known to the decoder, this similarity provides a means to maintain synchronization in the presence of random insertions and deletions. Substitution errors which remain after achieving synchronization are dealt with through regular error correction, making use of available redundancy.

Although watermark barcodes appear promising, their practical design for sequencing applications remains an open problem: for SMRT-CLR sequencing error rates, even the best barcodes reported in \cite{Kracht} yield sample misassignment rates in the vicinity of 5\%. In this paper, we revisit the design of DNA barcodes around the watermark concept. As opposed to \cite{Kracht}, we consider short low-density parity check (LDPC) codes \cite{mackay1999} as outer codes, which offer the interesting possibility of discarding very noisy reads rather than risk erroneous decoding. In addition, as our main contribution, the key watermark-carrier imprinting that conveys resilience to challenging insertion and deletion errors is modified so that non-sparse carriers are now allowed. This is accomplished by introducing a non-sparse inner code and an appropriate decoding algorithm built upon an adaptation of \cite{Briffa}, and is shown to significantly improve multiplexing performance. Further, we propose an algorithm that leverages knowledge about the chemical context where the barcodes are embedded to detect their boundaries. Finally, we show that the number of barcodes which are chemically suitable for use on the sequencing platform can be increased by exploiting the arbitrariness of the watermark. Together, these design enhancements define the flexible class of non-sparse watermark~(NS-watermark) barcodes, which offer high multiplexing capacity and are sufficiently robust for use in SMRT-CLR sequencing applications.
\section*{Results}
\begin{figure}[h!]
\centering
\includegraphics[scale=1.5]{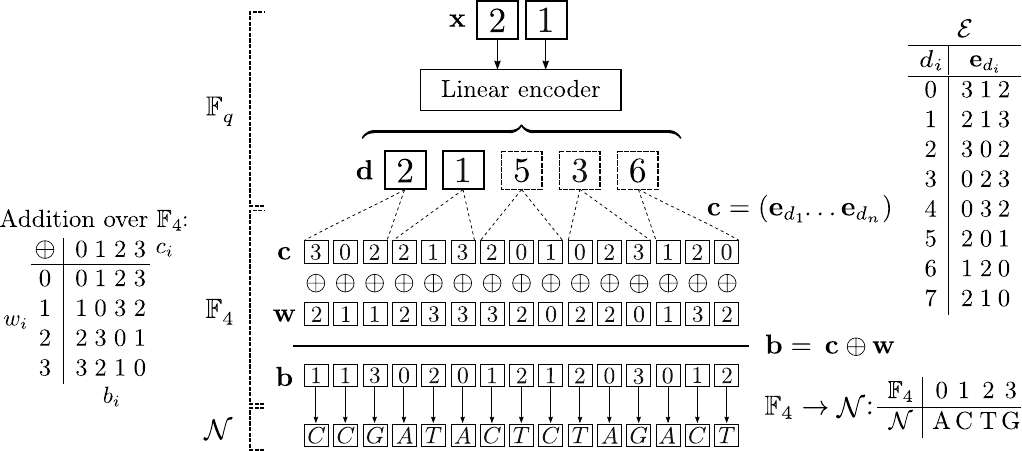}
\caption{\textbf{Construction of a NS-watermark barcode}, illustrated for the message $\mathbf{x} = 21 \in \mathbb{F}_8^2$, representing a particular sample number, and for parameters $q = 8$, $n = 5$, $k = 2$, $m = 3$ and $u = 3$. The specific inner codebook $\mathcal{E}$ and mapping $\mathbb{F}_4 \rightarrow \mathcal{N}$ used are shown on the right. The definition of addition over $\mathbb{F}_4$ is shown on the left. The linear encoder is assumed to enforce the following linear constraints: $d_1 = x_1$; $d_2 = x_2$; $d_3 = 2 x_1 + x_2$; $d_4 = x_1 + x_2$; $d_5 = 2 x_1 + 2 x_2$.}
\label{fig:barcodeConstructionGraphic}
\end{figure}

\begin{figure}[h!]
\centering
\includegraphics[scale=1]{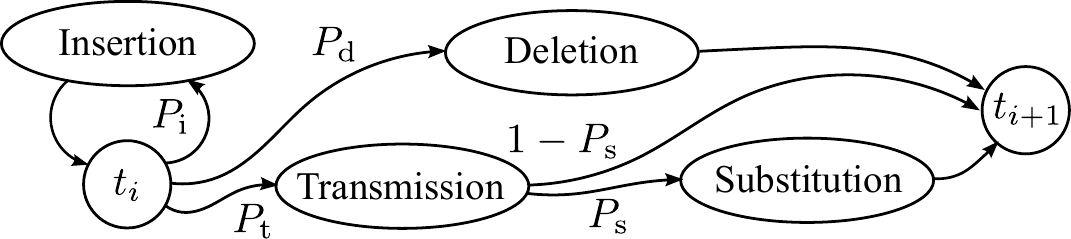}
\caption{\textbf{IDS model for errors affecting individual bases} $t_i$ during the sequencing process.}
\label{fig:channelModel}
\end{figure}

To allow correct demultiplexing in the presence of errors, redundancy must be added to sequencing barcodes. While this can be done directly in the domain where errors naturally occur, it has been shown that better performance can be achieved by ``packing'' low-level symbols together and designing codes in higher order finite fields \cite{LDPCoverGFq}. A finite field of order $q$, denoted by $\mathbb{F}_q$, is an alphabet of $q$ symbols with special rules for addition, subtraction, multiplication and division. Watermark barcodes, introduced in \cite{Kracht} and revisited here, exploit the above fact and embed redundancy into an information message through a combination of an outer code, defined on a high order finite field $\mathbb{F}_q$, and an inner code, which operates at the level of nucleotides or quaternary ($\mathbb{F}_4$) symbols.
\paragraph*{Watermark barcodes based on short LDPC outer codes}
The outer code adds redundancy to an information message, which encodes the sample number, to protect it against substitution errors. This message is represented as a sequence $\mathbf{x}$ of length $k$ whose elements belong to $\mathbb{F}_q$, i.e. $\mathbf{x} \in \mathbb{F}_q^k$. Redundancy is introduced by a linear error correcting code, which encodes each of the $q^k$ possible values of $\mathbf{x}$ into an ``outer codeword'' $\mathbf{d}  \in \mathbb{F}_q^n$ of length $n$, that carries $k$ informative symbols and $m \coloneqq n - k$ redundant ones (Fig.~\ref{fig:barcodeConstructionGraphic}). In a linear code, redundancy is added in such a way that the elements $d_i$ of $\mathbf{d}$ satisfy a series of linear constraints (e.g. $d_4 = 3d_1 + 4d_2$). In \cite{Kracht}, tabulated linear codes found by exhaustive methods and collected in \cite{MAGMA} were used as outer codes. Instead, we preserve the original formulation of watermark codes relying on powerful low-density parity check~(LDPC) codes \cite{mackay1999}. In particular, we use short quaternary LDPC codes developed in \cite{Tapia2015} for DNA barcoding applications affected mainly by substitution errors. These can be easily extended to arbitrary order fields, unlike the codes collected in \cite{MAGMA}, which are currently limited to $\mathbb{F}_9$.
\paragraph*{Watermark barcodes based on non-sparse inner codes}
Given an outer codeword $\textbf{d}$, the inner code expands each symbol $d_i \in \mathbb{F}_q$ into a quaternary sequence $\mathbf{e}_{d_i}$ of fixed length $u$, taken from an inner codebook $\mathcal{E}$ of size $q \times u$. As a result of this expansion, quaternary carriers $\mathbf{c}$ of length $l \coloneqq n u$ are obtained. Carriers are then imprinted with a known quaternary watermark sequence $\mathbf{w}$ of the same length (by simple symbol-wise addition over $\mathbb{F}_4$), resulting in imprinted sequences $\mathbf{b}$. Finally, using a fixed mapping from $\mathbb{F}_4$ into the ``nucleotide space'' $\mathcal{N} \coloneqq \{ A,C,T,G \}$, a set of $q^k$ candidate barcode sequences is obtained (Fig.~\ref{fig:barcodeConstructionGraphic}). In \cite{watermarks} and \cite{Kracht}, the sequences of $\mathcal{E}$ ($\mathbf{e}_0$ through $\mathbf{e}_{q-1}$) are constrained to be sparse. This helps achieve synchronization by introducing relatively few modifications to the watermark. However, it is accomplished at the expense of increased similarity between watermark-imprinted codewords, which may in turn lead to diminished error correction performance. Still, if codewords are sufficiently long (e.g., with hundreds or thousands of symbols, as in communication applications), the blessing face of dimensionality turns such effect unnoticeable. Conversely, if the length of codewords is reduced to the range of tens of symbols, as in the case of DNA barcodes, similarity between codewords may become a major concern, particularly at high levels of noise, as in the case of SMRT-CLR sequencing. This may explain the poor performance of watermark barcodes recently proposed in \cite{Kracht} for this range of error rates. To overcome this problem, we instead select the sequences of $\mathcal{E}$ with no constraints other than large pairwise hamming distance (i.e. choosing them to be as different as possible). As a result, the minimum edit distance factor, which is key for the performance of any coding scheme, improves significantly.
\paragraph*{Barcode filtering}
Of the initial $M \coloneqq q^k$ candidate barcode sequences, only a reduced number $B$ will be chemically suitable for the SMRT sequencing platform. Different factors can reduce multiplexing capacity, including GC content, homopolymers, primer dimer formation and compatibility with sequencing adapters. In order to account for these effects and filter out unsuitable barcodes, we consider a filtering stage based on a version of the \texttt{barcrawl} filtering tool \cite{barcrawl} adapted to accept external candidate barcodes. Specifically, barcodes are filtered based on the following criteria: GC content between 35\% and 65\%, maximum homopolymer length of 5, maximum heteroduplex length of 6 and maximum hairpin length of 6. This filtering stage considers not only the individual barcodes but also their compatibility with one another and with the SMRT sequencing adapter.
\paragraph*{On the choice of the watermark sequence}
As long as it is fixed and known, the watermark can be arbitrarily chosen, and a random sequence is the usual choice \cite{watermarks, Kracht}. However, random watermarks can perform poorly when it comes to satisfying chemical constraints. To overcome this limitation, we propose a method that exploits the arbitrariness of the watermark string to minimize the loss of multiplexing capacity due to chemical constraints. More precisely, we traverse the watermark from left to right and, at each position, select the base which maximizes the number of surviving barcodes after the \texttt{barcrawl} filtering stage. This process is repeated until no improvement is achieved in an entire pass (indicating a local minimum). At this point, a fixed number of bases are changed at random before traversing the pattern again. In essence, this is an iterated local search \cite{gendreau2010} where the number of barcodes lost during the filtering stage is taken as the cost function. For the seven NS-watermark barcode sets discussed later in this paper, this simple heuristic approach yields an average {\raise.17ex\hbox{$\scriptstyle\sim$}}1.5-fold increase in multiplexing capacity relative to the use of random watermarks, with no apparent degradation in multiplexing performance.
\paragraph*{A model for SMRT sequencing errors}
In order to formally describe indels and substitutions introduced by the sequencing process, DNA sequences were modeled as being transmitted over a noisy channel. For this purpose, the Insertion Deletion Substitution (IDS) channel model, defined in \cite{watermarks} and adapted to sequencing in \cite{Kracht}, was considered (Fig.~\ref{fig:channelModel}). For each incoming base, a random base can be inserted with probability $P_\mathrm{i}$ (which can happen a maximum of $I$ times), and then the current base is either deleted with probability $P_\mathrm{d}$ or sequenced (``transmitted'') with probability $P_\mathrm{t}$. Sequenced bases can themselves be correctly sequenced with probability $1 - P_\mathrm{s}$ or suffer a substitution error with probability $P_\mathrm{s}$. When a substitution occurs, the three possible base replacements are equiprobable.
\subsection*{Demultiplexing NS-watermark barcodes}
\begin{figure}[h!]
\centering
\includegraphics[scale=1.5]{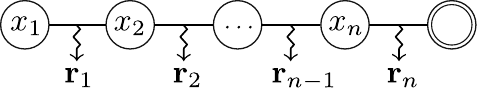}
\caption{\textbf{HMM $\mathcal{M}$ for inner decoding.} The double circle represents a boundary condition.}
\label{fig:HMM}
\end{figure}

To demultiplex corrupted watermark barcodes, a two-step decoding process is used. An inner decoder first processes the raw received sequence, affected by indels and substitutions, and produces an outer codeword corrupted only by probabilistic substitution errors. This codeword then enters an outer decoder which recovers the original information, i.e. the sample number.
\paragraph*{Inner decoding}
Let $\mathbf{r}$ be the sequence of quaternary symbols obtained by sequencing a NS-watermark barcode and mapping it back from $\mathcal{N}$ to $\mathbb{F}_4$. We know that $\mathbf{r}$ ultimately comes from an outer codeword $\mathbf{d}$ whose symbols $d_1 \ldots d_n$ have been expanded into sequences of size $u$ and watermarked before being sequenced. Based on this, let $\mathbf{r}_i$ be the sub-sequence of $\mathbf{r}$ that corresponds to $d_i$. If no indels occur, $\mathbf{r}_i=\left[r_{(i-1)u+1} \ldots r_{iu}\right]$. If we do admit indels and define the drift $x_i$ at the start of the transmission of $d_i$ as the difference between insertions and deletions up to that point, then $\mathbf{r}_i=[r_{(i-1)u + 1 + x_i} \ldots r_{i u + x_{(i+1)}}]$. Since the probability of an indel occurring does not depend on errors made in the past, the Markov property $P(x_{i+1} | x_i \ldots x_1) = P(x_{i+1} | x_i)$ holds for drift variables. Therefore, the process of sequencing expanded and watermarked codeword symbols $d_i$ can be modeled as a Hidden Markov Model (HMM) $\mathcal{M}$ of the Mealy type (with emissions on transitions), with drifts $x_i$ as hidden states and sub-sequences $\mathbf{r}_i$ as observables (Fig.~\ref{fig:HMM}).

Given $\mathcal{M}$, the well-known forward-backward (FB) algorithm can be applied to calculate the likelihoods $L(d_i = a) \coloneqq P(\mathbf{r} | d_i = a)$ for $i = 1 \ldots n$ and $a = 0 \ldots q-1$, which will initialize the outer decoder:
\begin{equation}\label{eq:likelihood1}
L(d_i = a) = \mathlarger{\sum}_{x^-, x^+} \underbrace{P(\mathbf{r}^-, x_{i} = x^{-})}_{F_{i} (x^{-})} \underbrace{P(\mathbf{r}_i, x_{i+1} = x^{+} | d_i = a, x_{i} = x^{-})}_{P(\mathbf{r}_i, x^{-} \rightarrow x^{+} | d_i = a)} \underbrace{P(\mathbf{r}^+ | x_{i+1}=x^{+})}_{B_{i+1}(x^{+})}\text{,}
\end{equation}
where $\mathbf{r}^- \coloneqq \left[ \mathbf{r}_{1} \dots \mathbf{r}_{i-1} \right]$ and $\mathbf{r}^+ \coloneqq \left[\mathbf{r}_{i+1} \dots \mathbf{r}_{n}\right]$. In \eqref{eq:likelihood1}, $F_{i} (x^{-})$ and $B_{i+1}(x^{+})$ are known as forward and backward quantities, respectively, and can be computed from $\mathcal{M}$ using the standard formulation of the FB algorithm. The calculation of forward quantities is made recursively from an initial boundary condition $F_1(x)$ (representing \textit{a priori} knowledge about the initial drift) and is known as a ``forward pass''. Similarly, the recursive calculation of backward quantities from a final boundary condition $B_{n+1}(x)$ (representing \textit{a priori} knowledge about the final drift) is known as a ``backward pass''. Lastly, $P(\mathbf{r}_i, x^{-} \rightarrow x^{+} | d_i = a)$ is the probability that drift changes from $x^{-}$ to $x^{+}$ during the transmission of $d_i$ and such transmission results in the reception of $\mathbf{r}_i$, given that $d_i$ is equal to $a$. This value requires an additional stage of computation using a ``nucleotide-level'' HMM $\mathcal{H}$, which is analogous to $\mathcal{M}$ but operates at the level of individual quaternary symbols or nucleotides. A more detailed discussion of the computation of \eqref{eq:likelihood1} can be found in the Appendix.
\paragraph*{Outer decoding}
By computing the likelihoods $L(d_i)$ (i.e. the likelihood that each symbol of $\mathbf{d}$ took each of the $q$ possible values), the inner decoder effectively transforms any combination of indels and substitution errors affecting $\mathbf{r}$ into probabilistic substitution errors, for which the outer linear code was specifically designed. Given these likelihoods, generic linear codes can be decoded using Maximum-Likelihood~(ML) approaches, which select the codeword that maximizes the probability of receiving $\mathbf{r}$. While this is mathematically optimal, there always exists a codeword which maximizes such probability and, thus, decoding never fails. As noted in \cite{Tapia2015}, incomplete decoders (i.e. decoders that report a decoding failure when the result is ambiguous) can be used instead to control the trade-off between detected errors (read losses) and undetected errors (sample misassignments). In multiplex sequencing applications, it is usually far preferable to discard a read than to assign it to an incorrect sample. Additionally, ML decoding scales poorly \cite{wolf1978}, which becomes prohibitive as we explore longer barcodes and higher multiplexing capacities.

\begin{figure}[h!]
\centering
\includegraphics[scale=1.5]{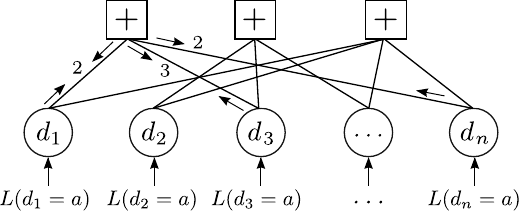}
\caption{\textbf{Illustration of outer decoding via Belief Propagation.} Edge weights and belief messages are shown only for the first constraint, which is $2 d_1 + 3 d_3 + 2 d_n = 0$ (with $\mathbb{F}_q$ arithmetic).}
\label{fig:messagePassing}
\end{figure}

In the particular case of LDPC codes, an iterative decoding algorithm known as Belief Propagation~(BP) \cite{mackay1999} can be used instead to simultaneously address both issues. Specifically, complexity now scales linearly with $n$ \cite{LDPCoverGFq}, while the maximum number of iterations can be used to control the trade-off between read losses and sample misassignments, as explained below. Although the details are outside the scope of this manuscript, BP can be intuitively understood as a message passing algorithm on a graph like that shown on Fig. \ref{fig:messagePassing}. The graph includes a set of variable nodes (labeled $d_1$ through $d_n$) which represent the $n$ symbols of an outer codeword, a set of constraint nodes (labeled $+$) and a set of connecting edges. A constraint node is said to be satisfied if the sum of variable nodes connected to it is zero (with addition defined over the corresponding finite field and each variable weighted by an appropriate constant). Connecting edges are drawn so that every constraint is simultaneously satisfied if and only if a set of values for $d_1$ through $d_n$ forms a valid codeword, thus providing a compact graphic representation of the code structure.
Decoding begins with an initial guess for the probability distribution of each variable node, given by $L(d_i = a)$, as calculated by the inner decoder. If the value with the highest probability is selected for each variable node (which is known as ``hard thresholding'') and the resulting set of values satisfies every constraint, a valid codeword has been found and decoding stops. If, however, at least one constraint is not satisfied, the BP algorithm draws progressively close to a valid codeword by iteratively exchanging belief ``messages'' between variable and check nodes. After a variable number of iterations, the algorithm is expected to converge to a distribution which, when hard thresholded, satisfies every constraint and thus corresponds to a valid codeword. If this does not occur within a predefined number of iterations, a decoding failure is reported. In our sequencing context, the latter means that the received sequence is too noisy to be confidently decoded and the read is simply discarded. If we admit few iterations, only barcodes with few errors will reach convergence and decoding will be very conservative (few sample misassignments but many discarded reads). Conversely, if we admit many iterations, few reads are discarded but more reads are misassigned.
\subsection*{Context-aware boundary estimation}
\begin{figure}[h!]
\centering
\includegraphics[scale=2.25]{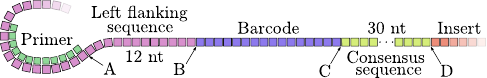}
\caption{\textbf{Sequencing adapter for a Barcoded Universal Primer experimental setup.}}
\label{fig:SMRTbell}
\end{figure}

The forward and backward passes in the inner decoder require boundary conditions given by $F_1(x)$ and $B_{n+1}(x)$, respectively. If the barcode were sequenced in isolation, then $F_1(x)$ would be $1$ for $x=0$ and $0$ otherwise (the drift at the start of the barcode would be necessarily $0$). Similarly, $B_{n+1}(x)$ would be $1$ for $x = \mathrm{len}(\mathbf{r}) - \mathrm{len}(\mathbf{b})$ and $0$ otherwise (the final drift would be known). In practice, however, a sequencing read also includes the sequenced insert and certain platform-specific sequences. For multiplex SMRT sequencing, a typical experimental setup is known as ``Barcoded Universal Primer'' \cite{barcodedUniversalPrimer}, where the insert is capped by a so-called SMRTbell\textsuperscript{TM} adapter \cite{CCS}, as shown in Fig. \ref{fig:SMRTbell}. In an error-free read, the barcode will be flanked to the left by a 12-nucleotide sequence located immediately after the primer annealing site and to the right by a 30-nucleotide consensus sequence, followed by the insert.
If indels occur, barcode boundaries may shift relative to their expected positions, and this must be taken into account for successful decoding. A simple approach to do this is to consider a uniform distribution for $F_1(x)$ and $B_{n+1}(x)$, as done in \cite{Kracht}. However, since the flanking sequences are known, a better result can be obtained by using these as synchronizing markers. Specifically, the left barcode boundary (B) is estimated by performing a nucleotide-level forward pass along the 12-nucleotide left flanking sequence. Because sequencing begins at this point, this forward pass can itself be initialized with a drift of 0 at the start of the left flanking sequence (A). Similarly, the right barcode boundary (C) is estimated by performing a nucleotide-level backward pass along the 30-nucleotide consensus sequence, which is initialized with a uniform distribution at the end of such sequence (D).
\subsection*{Simulation setup}
To assess their performance \textit{in silico}, existing watermark barcodes and the proposed NS-watermark alternative were flanked to the left and right by appropriate sequences to account for the sequencing adapter and transmitted through the IDS channel model described previously. The sequenced insert was represented by an indefinitely long random sequence. For codes reported in \cite{Kracht}, the boundary estimation and decoding algorithms reported by the authors were used. Specifically, forward and backward recursions were initialized with a uniform distribution and a soft maximum likelihood linear decoder \cite{wolf1978} was used as outer decoder. For NS-watermark codes, our context-aware boundary estimation algorithm, our inner decoder and a regular BP LDPC decoder with a maximum of 10 iterations were used. The sample misassignment probability $P_{u}$ was then estimated by Monte Carlo simulation. For each set of barcodes, $N = 5\times10^{7}$ barcode sequences were flanked, transmitted through the channel model, passed through the appropriate boundary detection algorithm and decoded. Barcodes were taken in equal numbers from the sets of filtered barcodes. The proportion ${\bar{P}}_\mathrm{u}$ of barcodes for which the decoder output was different from the original sample was computed and taken as an estimate of $P_\mathrm{u}$. If ${\bar{P}}_\mathrm{u} \neq 0$, then 95\% confidence intervals $\left[{{\bar{P}}_\mathrm{u}^-}, {{\bar{P}}_\mathrm{u}^+}\right]$ were computed as ${\bar{P}}_\mathrm{u}~\mathrm{exp}\left(\pm 2\sigma\right)$, where $\sigma = \sqrt{(1-{\bar{P}}_\mathrm{u})/(N {\bar{P}}_\mathrm{u})}$. If ${\bar{P}}_\mathrm{u}=0$, then ${\bar{P}}_\mathrm{u}^- = 0$ and ${\bar{P}}_\mathrm{u}^+=1-\mathrm{exp}\left(-2/N\right)$ were used instead \cite{mackay1999}. In the case of NS-watermarks, where the BP decoder is incomplete and may report a decoding failure, a similar procedure was used to estimate the read loss probability $P_\mathrm{e}$. The theoretical multiplexing capacity $M$ was calculated as $q^k$, while the actual number of barcodes $B$ was obtained by counting the barcodes that were compatible with SMRT chemical sequencing constraints (i.e. that survived filtering). All NS-watermark barcode sets considered in this manuscript are available at \url{www.cifasis-conicet.gov.ar/ezpeleta/NS-watermark}.
\subsection*{Multiplexing performance of watermark barcodes}
\begin{figure}[h!]
\centering
\hspace{-1cm}
\makebox[\textwidth][c]{\includegraphics[width=1.1\textwidth]{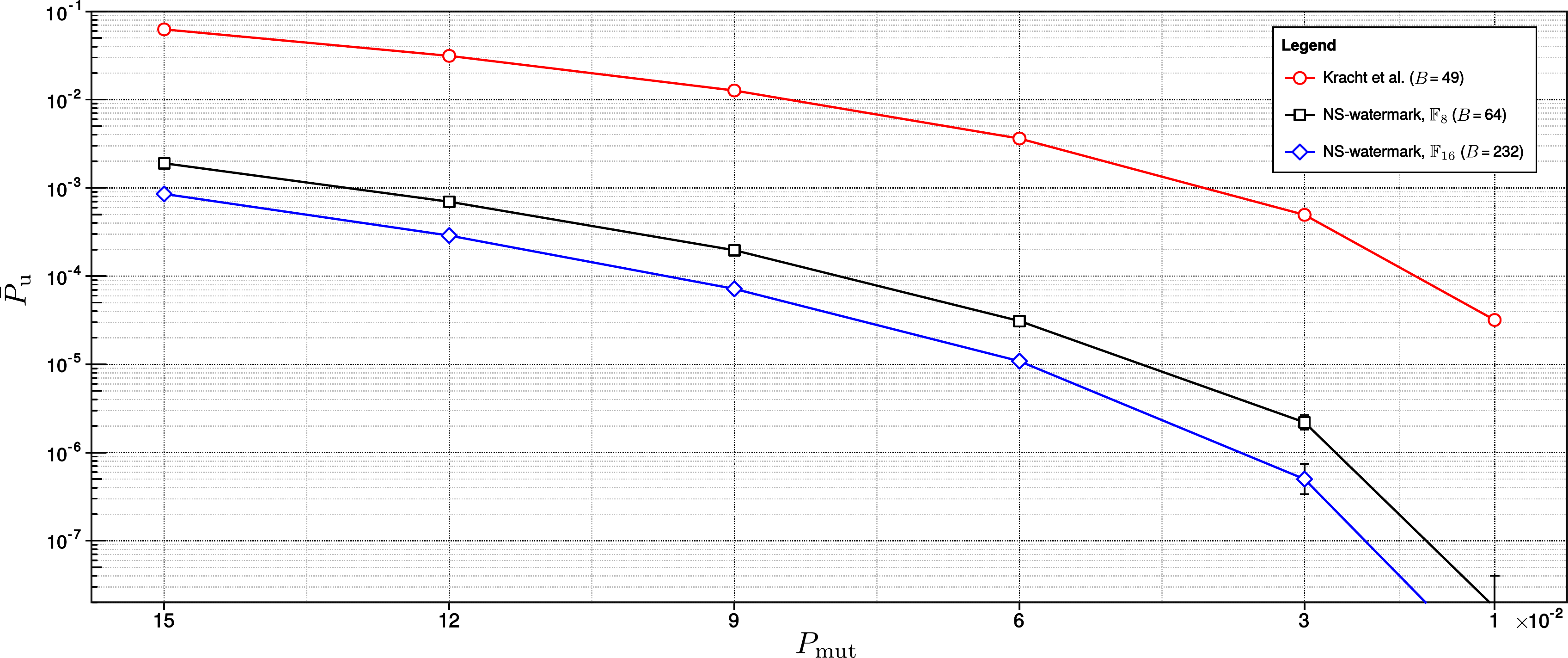}}
\caption{{\bf Sample misassignment rates for three types of 24-nucleotide barcodes as a function of the mutation probability $P_\mathrm{mut}$.}
The number of barcodes $B$ is shown in parenthesis in the legend. 95\% confidence intervals $\left[{{\bar{P}}_\mathrm{u}^-}, {{\bar{P}}_\mathrm{u}^+}\right]$ are shown as error bars where not negligible.}
\label{fig:resultsVsKracht}
\end{figure}

Simulation experiments were performed to compare the overall robustness of watermark barcodes recently introduced in \cite{Kracht} and that of the proposed NS-watermark barcoding alternative under different levels of sequencing errors (Fig.~\ref{fig:resultsVsKracht}). For this comparison, we selected the best codes reported in \cite{Kracht} in terms of error performance ($q=7$, $k=2$, $n=6$, $u=6$) and constructed NS-watermark barcode sets of the same length ($l = 24$) with outer codes defined over $\mathbb{F}_8$ and $\mathbb{F}_{16}$ (both with $k = 2$, $n = 6$, $u = 4$). For this simulation, mutation probabilities $P_\mathrm{mut} \in \left[ 0.01 \dots 0.15 \right]$ were considered, where $P_\mathrm{mut} \coloneqq P_\mathrm{i} + P_\mathrm{d} + P_\mathrm{s}$ and $P_\mathrm{i} = P_\mathrm{d} = P_\mathrm{s}$.
\subsection*{Multiplexing performance of NS-watermark barcodes for SMRT sequencing}
On a second simulation, the performance of NS-watermark barcodes of length $24$, $48$ and $96$ with outer codes defined over $\mathbb{F}_{16}$ was evaluated for particular error probabilities which are representative of the SMRT error profile~(Table~\ref{tableSMRT}). Specifically, we considered $P_\mathrm{i} = 0.055$, $P_\mathrm{d} = 0.055$ and $P_\mathrm{s} = 0.01$. For each barcode length, different values of $k$ (i.e. different multiplexing capacities) were also considered.
\begin{table}[!ht]
\centering
\begin{tabular*}{\textwidth}{@{\extracolsep{\fill}}c c c c c c c}
\hline
$l$ & $M$ & $B$ & $\bar{P}_\mathrm{e}$ & $\bar{P}_\mathrm{e}^+$ & $\bar{P}_\mathrm{u}$ & $\bar{P}_\mathrm{u}^+$ \\ \hline
24 & 256 & 232 & $7.2\times10^{-3}$ & $7.3\times10^{-3}$ & $1.7\times10^{-4}$ & $1.8\times10^{-4}$ \\{\vspace{5pt}}24 & 4096 & 3567 & $2.3\times10^{-2}$ & $2.3\times10^{-2}$ & $2.5\times10^{-3}$ & $2.5\times10^{-3}$ \\
48 & 256 & 239 & $7.0\times10^{-5}$ & $7.2\times10^{-5}$ & $2.2\times10^{-7}$ & $4.0\times10^{-7}$ \\{\vspace{5pt}}48 & 4096 & 3471 & $2.3\times10^{-4}$ & $2.4\times10^{-4}$ & $1.1\times10^{-6}$ & $1.4\times10^{-6}$ \\
96 & 256 & 164 & $9.8\times10^{-6}$ & $1.0\times10^{-5}$ & $0$ & $4.0\times10^{-8}$ \\
96 & 4096 & 2163 & $1.0\times10^{-5}$ & $1.1\times10^{-5}$ & $0$ & $4.0\times10^{-8}$ \\
96 & 65536 & 32136 & $1.9\times10^{-5}$ & $2.0\times10^{-5}$ & $6.2\times10^{-8}$ & $2.0\times10^{-7}$ \\ \hline
\end{tabular*}
\caption{\small {\bf Performance of NS-watermark barcode sets ($\mathbb{F}_{16}$) under the SMRT error profile} for varying length $l$ and unfiltered multiplexing capacity $M$. $B$ is the number of barcodes, $\bar{P}_\mathrm{e}$ is the read loss probability and $\bar{P}_\mathrm{u}$ is the sample misassignment probability. $\bar{P}_\mathrm{e}^+$ and $\bar{P}_\mathrm{u}^+$ are upper error bars for $\bar{P}_\mathrm{e}$ and $\bar{P}_\mathrm{u}$.}
\label{tableSMRT}
\end{table}
\section*{Discussion}
As seen in Fig. \ref{fig:resultsVsKracht}, NS-watermark barcodes consistently outperform the best watermark barcodes reported in \cite{Kracht} for the same barcode length ($l = 24$), while simultaneously increasing multiplexing capacity. Additionally, results show that both the decoding failure probability $P_\mathrm{e}$ (discarded reads) and the undetected demultiplexing error probability $P_\mathrm{u}$ (misassigned reads) decrease monotonically with read length (Table \ref{tableSMRT}). In connection with this, we note that, although working with CLRs introduces significantly higher error rates than using CCSs, it also makes it possible to use longer barcodes while keeping the relative barcoding overhead within reasonable limits. For example, $96$~bp barcodes, which are unfeasible for CCS reads or short read technologies, introduce a relative overhead of less than 1\% on the average CLR of around 10~kb.  

We further note that, because of the low sample misassignment rates, correctly assigned reads are expected to vastly outnumber misassigned reads for any given sample. Under these conditions, misassigned reads are likely to be ``washed away'' by downstream consensus within each sample group and, therefore, the actual per-base error due to undetected demultiplexing errors could be several orders of magnitude smaller than the reported sample misassignment rate.

A major advantage of the proposed NS-watermark is the flexibility it offers for code construction. Within reasonable ranges, any combination of $u$, $q$, $n$, $l$, $k$, $m$, $\mathbf{w}$ and $\mathcal{E}$ yields an admissible barcode set. This increased flexibility, along with the systematic construction method, means new code configurations can be explored virtually effortlessly to adapt to changing requirements, given by admissible rates of read losses and sample misassignments, acceptable coding overhead, number of samples, or even error profiles. In connection with the latter, our design method can be easily extended to other third generation sequencing technologies impaired by high rates of indels and sequencing errors.
\bibliographystyle{plain}
{\normalfont

}
\section*{Appendix}
During discussion of inner decoding, it was mentioned that the likelihoods $L(d_i)$, needed by the outer decoder, could be computed as
\begin{equation}\label{eq:likelihood2}
L(d_i = a) = \sum_{x^-, x^+ \in X} F_{i}(x^{-}) P(\mathbf{r}_i, x^{-} \rightarrow x^{+} | d_i = a) B_{i+1}(x^{+})\text{.}
\end{equation}
It was mentioned that $F_{i} (x^{-})$ and $B_{i+1}(x^{+})$ could be computed from $\mathcal{M}$ using forward and backward recursions. More precisely:

\begin{equation}\label{eq:symbolLevelForwardQuantity}
F_j(x_f) = \mathlarger{\sum}_{\substack{x^-_f \in X\\ a \in \mathbb{F}_q}} F_{j-1}(x^-_f) P(\mathbf{r}_{j-1}, x^-_f \rightarrow x_f | d_{j-1} = a)\text{~and} 
\end{equation}
\begin{equation}\label{eq:symbolLevelBackwardQuantity}
B_k(x_b) = \mathlarger{\sum}_{\substack{x^+_b \in X\\ a \in \mathbb{F}_q}} B_{k+1}(x^+_b) P(\mathbf{r}_{k}, x_b \rightarrow x^+_b | d_{k} = a)\text{,} 
\end{equation}
where $X$ is the space of all possible drift values. In \eqref{eq:likelihood2}, \eqref{eq:symbolLevelForwardQuantity} and \eqref{eq:symbolLevelBackwardQuantity}, terms of the form $P(\mathbf{r}_{s}, x_1 \rightarrow x_2 | d_{s} = a)$ represent the probability that drift changes from $x_1$ to $x_2$ during the transmission of $d_s$ and such transmission results in the reception of $\mathbf{r}_s$, given that $d_s$ is equal to $a$. To compute these quantities, let $\mathbf{t}_s$ be the transmitted sub-string that would correspond to $d_s$ if $d_s$ were indeed equal to $a$ (i.e. $\mathbf{e}_{a}$ plus the appropriate symbols of $\mathbf{w}$, namely $w_{(s-1)u + 1} \dots w_{su}$). Further, note that the probability of receiving $\mathbf{r}_{s}$ and the drift changing from $x_1$ to $x_2$ is equal to the probability of receiving $\mathbf{r}_{s}$ and the drift changing \textit{by} $\Delta x \coloneqq x_2-x_1$, since the probability of new indels does not depend on the current drift. Moreover, the drift change $\Delta x$ is implicit in the lengths of $\mathbf{r}_s$ and $\mathbf{t}_s$, since $\Delta x = \mathrm{len}(\mathbf{r}_s) - \mathrm{len}(\mathbf{t}_s)$. In light of the above, $P(\mathbf{r}_{s}, x_1 \rightarrow x_2 | d_{s} = a)$ can be simplified to $P(\mathbf{r}_{s} | \mathbf{t}_{s})$.

\begin{figure}[h!]
\centering
\includegraphics[scale=1.5]{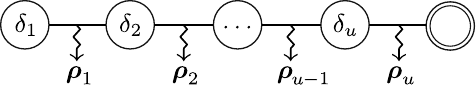}
\caption{\textbf{Nucleotide-level HMM $\mathcal{H}$.} The double circle represents a boundary condition.}
\label{fig:HMM_quad}
\end{figure}

To calculate $P(\mathbf{r}_{s} | \mathbf{t}_{s})$, we introduce a new HMM $\mathcal{H}$~(Fig.~\ref{fig:HMM_quad}), which is analogous to $\mathcal{M}$ except that hidden variables $\delta_1 \dots \delta_u$ now represent the drift before transmitting each individual nucleotide and observables $\boldsymbol\rho_1 \dots \boldsymbol\rho_u$ represent (possibly empty) strings received from the transmission of a single nucleotide through the IDS channel model~(Fig.~\ref{fig:channelModel}).

\noindent $P(\mathbf{r}_s | \mathbf{t}_s)$ is then calculated by performing a nucleotide-level forward pass over the $u$ states of $\mathcal{H}$, according to \eqref{eq:receiverLikelihood} and \eqref{eq:qitLevelForwardQuantity} below:
\begin{equation}\label{eq:receiverLikelihood}
P(\mathbf{r}_s | \mathbf{t}_s) = f_{u+1}(\mathrm{len}(\mathbf{r}_s)-\mathrm{len}(\mathbf{t}_s))
\end{equation}
\begin{equation}\label{eq:qitLevelForwardQuantity}
f_y(\delta) = \mathlarger{\sum}_{\delta^- \in (\Delta \cap \{ \delta - I, \dots, \delta+1 \} )} f_{y-1}(\delta^-) P(\boldsymbol\rho_{y-1} | t_{y-1})\text{,}
\end{equation}
where $t_{y-1}$ is the $(y-1)$-th quaternary symbol of $\mathbf{t}_s$ and $P(\boldsymbol\rho_{y-1} | t_{y-1})$ is the probability of receiving the (possibly empty) string \mbox{$\boldsymbol\rho_{y-1} \coloneqq \left[ r_{s_{(y - 1 + \delta^-)}} \dots r_{s_{(y - 1 + \delta)}} \right]$} from the transmission of $t_{y-1}$, which is calculated according to \eqref{eq:emissionProbability} below. The boundary conditions for \eqref{eq:qitLevelForwardQuantity} are $f_1(\delta) = 1$ for $\delta = 0$, and $f_1(\delta) = 0$ for any other $\delta$, because the local drift before transmitting any symbols is necessarily $0$. A nucleotide-level backward pass, needed for the boundary estimation algorithm, is handled analogously to \eqref{eq:receiverLikelihood}.

The emission probability $P(\boldsymbol\rho^* | t^*)$ of receiving a sub-string $\boldsymbol\rho^*$ from the transmission of a single quaternary symbol $t^*$ can be obtained by inspection of the IDS channel model~(Fig.~\ref{fig:channelModel}):
\begin{equation}\label{eq:emissionProbability}
P(\boldsymbol\rho^* | t^*) = \left\{ \begin{array}{rl}
P_\mathrm{d} &\mbox{ if $\mu = 0$} \\
\left( \frac{P_\mathrm{i}}{4} \right)^{\mu} P_\mathrm{d} + \left( \frac{P_\mathrm{i}}{4} \right)^{\mu-1} \frac{1}{3} P_\mathrm{t} P_\mathrm{s} & \mbox{ if $1 \le \mu < I+1$, $\rho^*_\mu \ne t^*$} \\
\left( \frac{P_\mathrm{i}}{4} \right)^{\mu} P_\mathrm{d} + \left( \frac{P_\mathrm{i}}{4} \right)^{\mu-1} P_\mathrm{t} (1-P_\mathrm{s}) & \mbox{ if $1 \le \mu < I+1$, $\rho^*_\mu = t^*$} \\
\left( \frac{P_\mathrm{i}}{4} \right)^I  \frac{1}{3} (1 - P_\mathrm{d}) P_\mathrm{s} & \mbox{ if $\mu = I+1$, $\rho^*_\mu \ne t^*$} \\
\left( \frac{P_\mathrm{i}}{4} \right)^I (1 - P_\mathrm{d}) (1-P_\mathrm{s}) & \mbox{ if $\mu = I+1$, $\rho^*_\mu = t^*$} \\
0 & \mbox{otherwise} \\
\end{array} \right. \text{,}
\end{equation}
where $\mu \coloneqq \mathrm{len}(\boldsymbol\rho^*)$.

The summations on \eqref{eq:likelihood2}, \eqref{eq:symbolLevelForwardQuantity} and \eqref{eq:symbolLevelBackwardQuantity} should in principle iterate over the space of all possible drift values, which is computationally impractical. To reduce complexity, the iteration is limited to an ``outer drift space'' $X \coloneqq \lbrace x_{min}, x_{max} \rbrace$, which equates to considering a maximum drift of $x_{max}$ and a minimum drift of $x_{min}$ throughout the transmission of the current barcode. Similarly, when calculating nucleotide-level passes according to \eqref{eq:qitLevelForwardQuantity}, an ``inner drift space'' $\Delta \coloneqq \lbrace \delta_{min}, \delta_{max} \rbrace$ is considered, which equates to considering a maximum local drift of $\delta_{max}$ and a minimum local drift of $\delta_{min}$ during the transmission of a single outer symbol.
\end{document}